# Cavity Growth in Soft Solids

Arnaud Chiche, Josef Dollhofer and Costantino Creton*
*Laboratoire PCSM - ESPCI, 10, Rue Vauquelin, 75231 Paris Cedex 05, France*

Soft polymer-based solid materials can be defined as solid materials with elastic moduli below 1 MPa. In order to obtain such a low modulus while retaining a solid character, all these materials are based on a network of fixed points (crosslinks, physical interactions) keeping together a liquid-like matrix. Since they are not able to flow, very large deformations will lead to failure by fracture or cavitation. We discuss in this paper new experimental data suggesting that the growth of a cavity in a soft nearly elastic solid undergoes a transition from a stress-controlled growth ($\mathbf{s} > \mathbf{s}_{cav}$), when the initial cavity size is larger than $G_c/E$, to an energy-activated growth, when the initial cavity size is smaller than $G_c/E$. The important material length scale $G_c/E$ represents the ratio between the surface energy necessary to expand the cavity and the elastic modulus. This energy activated growth regime is due to the existence of a metastable solution and significantly increases the apparent cavitation stress for layers thinner than 1000 $G_c/E$.

Despite significant industrial interest in the topic, the criteria triggering fracture of soft materials, as commonly encountered in many applications such as self-adhesive products, cosmetics or biological tissues [1], have remained much more elusive than in the case of brittle solids. In brittle solids the common view of failure is that the material has intrinsic flaws. Failure in a structure occurs by the growth of one of these flaws into a crack when the energy release rate $G$ (the elastic energy stored in the material which would be released by the growth of the flaw) reaches a critical value $G_c$, controlled by the dissipative processes occurring during crack propagation [2]. In this picture, only the largest flaw matters since it will propagate catastrophically and the criterion for propagation is that of a critical energy [3].

In soft materials of course intrinsic flaws exist as well, however their growth under an applied remote stress does not necessarily lead to a catastrophic failure of the material. The propagation of a single crack accompanied by the release of elastic energy is seldom observed. Failure occurs more often by a mechanism more analogous to tearing, where highly dissipative processes involving large scale deformations of the material occur. These processes invariably involve, locally, the formation of cavities and the orientation and extension of polymer chains, reinforcing dramatically the resistance to fracture [4-6].

The determination of the criteria controlling the formation of cavities in a soft material which are the focus of our experiments, become then essential to understand the resistance to fracture of this class of materials.

Cavitation, i.e. the nucleation and growth of voids under a tensile stress, has been investigated extensively both in simple liquids [7-9] and in elastic rubbers [10] (elastic moduli ~ 1MPa). Cavitation in liquids invariably occurs on preexisting defects [9] (heterogeneous nucleation) so that the measured cavitation stress is highly dependent on the defect population. Liquids cannot store elastic stress, so cavitation is controlled by a balance between viscous dissipation and surface tension. Cavitation in rubbers on the other hand, appears to be a process controlled chiefly by material properties, and in particular by the elastic modulus of the rubber [10]. Rubbers can store and release elastic energy and the growth of a cavity is always accompanied by the release of some elastic energy. Although the presence of defects is known, the level of stress at which cavities expand can be well predicted by the elastic properties of the rubber [11].

In between stiff rubbers and Newtonian fluids, exist a variety of polymeric-based soft materials with intermediate properties in terms of elastic modulus and viscoelastic character, where cavitation is observed [12-14], but appropriate criteria triggering it are largely unknown. We contribute here to this question by presenting new results focusing on elastic materials with an elastic modulus distinctly lower than that of a rubber.

Experimentally we brought in contact, and subsequently removed at a constant velocity, a flat-ended cylindrical probe (6 mm in diameter) from a thin (~100 μm) layer of soft block copolymer based adhesive gel. The experimental geometry is very well suited to provide a nearly hydrostatic stress to the material [15] and our adhesive gel is an ideal model material for these investigations. It is a styrene-isoprene-styrene block copolymer blended with small molecules which has a morphology of 15 nm styrene spherical domains dispersed in a rubbery matrix. This physically crosslinked gel sticks to a surface by simple contact because of its low elastic modulus around 0.2 MPa [16,17] and, while displaying some strain rate dependent properties, is predominantly elastic [5,18]. We show on figure 1, its linear viscoelastic and large strain nonlinear elastic properties at three different strain rates.

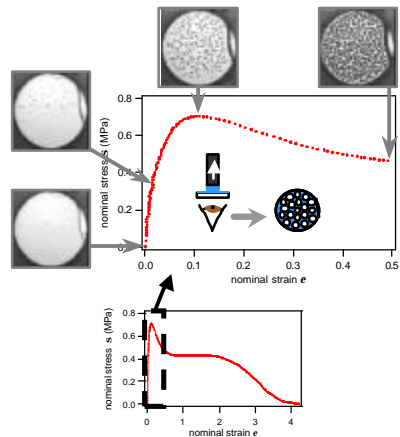

Fig. 2. Schematic of a probe test showing the nominal stress vs. nominal strain curve obtained from the test as well as characteristic images of cavities formed in the soft adhesive layer before the peak stress.

Our testing device [19] allows to measure precisely force and displacement during the tensile phase, while observing optically the growth of individual cavities (figure 2). The first peak in stress, referred to as $s_{max}$ in the following, corresponds broadly to the large-scale appearance of cavities while the plateau in stress corresponds to the formation of a stretched foam of walls between cavities which eventually detach cleanly from the surface. As shown on figure 2, after the peak stress, the layer dissipates a substantial amount of energy and cannot be considered elastic. We focus therefore our analysis on the initial part of the tensile curve. Presumably, the macroscopic cavities of figure 2 appeared where a small pre existing bubble was originally present at the interface. When the probe is brought in contact with the adhesive, the elastic properties of the adhesive do not allow an intimate contact everywhere at the interface so that a rough probe or a rough adhesive surface are able to create contact defects of average sizes depending on the topography of the two surfaces in contact [6,20]. In order to compare two different populations of initial cavities and to obtain reproducible results, we present in this paper experiments with two different adhesive/adherent interfaces. The first one consist of a nominally smooth adhesive surface in contact with a polished metal probe (rms roughness of 11 nm) [20]. In this case the size of the initial cavities is controlled by the roughness of the probe and they are not optically visible. The second one consists of the same probe in contact with a very rough adhesive layer containing holes at the surface. Once in contact with the probe, these holes form optically visible cavities.

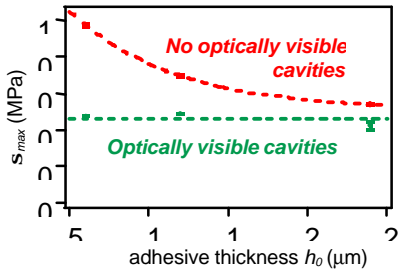

Fig. 3. Average values of the peak stress as a function of adhesive thickness. Red symbols: no optically visible cavities; green symbols: optically visible cavities.

The initial portion of representative tensile curves in this geometry showing the nominal stress (force normalized by initial contact area) vs. displacement of the probe are shown on figure 3 for three thicknesses of the adhesive layer containing two different populations of initial defects (cavities). When optically visible cavities are present at the interface at the beginning of the test (figure 3a), the first peak stress is less, if at all, visible and appears to be independent of the thickness of the layer. On the contrary when no initial cavities are optically visible, the value of $s_{max}$ increases sharply with decreasing layer thickness as shown on figure 3b. This increasing value of $s_{max}$ with decreasing layer thickness, shown on figure 3c as an average of at least five separate tests for each thickness, is surprising since it implies that the same interfacial defects expand at different levels of stress for different layer thicknesses.

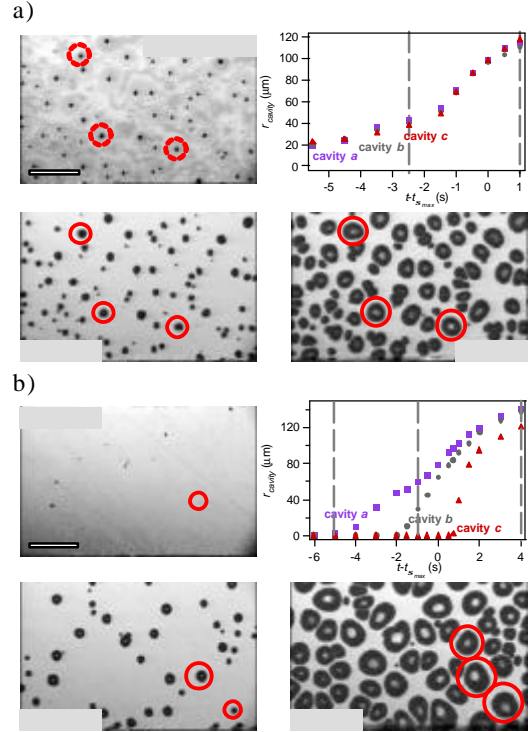

Figure 4: Representative images of the cavities at the interface between the probe and the film, and radius of three selected cavities of almost the same final size as a function of time during the debonding ($h_0$ = 120 µm, $V_{deb}$ = 10 µm.s$^{-1}$). $t_{s\,max}$ corresponds to the peak stress of figure 2. a) Large cavity case. The nearly simultaneous growth of all the cavities from large cavities initially present is evident. b) Small cavities case. The sequential appearance of new cavities with time is shown on these three images.

We observe that for the large defects case, all the observed cavities grow simultaneously from optically visible existing cavities until they reach a size of the order of the thickness of the layer (figure 4a). On the other hand, in the absence of optically visible defects, the cavities do not grow simultaneously but new cavities seem to appear independently and continuously with time. Once a cavity has appeared, it does grow to a size of the order of the thickness of the film but then does not grow any further. A substantial drop in stress only occurs when a sufficiently large number of these cavities have appeared. More importantly, the growth rate of each cavity depends on the level of stress at which it was nucleated: The growth of the first cavities is slow and progressive, while the growth of cavities appearing later is increasingly rapid (figure 4b).

When comparing the results for different layer thicknesses, our experiments show that for 240 µm thick layers (see figure 3c), the peak stress is barely affected by the presence or absence of optically visible defects. Following these observations, we can now examine in more detail the statistics of cavity appearance. Figure 5a shows the number of visible cavities per unit area as a function of applied stress for the case where no optically visible cavities are initially present. The analysis is limited to less than 10 cavities per square mm to remain in the macroscopic nearly elastic regime. From figure 5a, a nominally equivalent population of initial interfacial defects expands into a population of optically visible cavities at widely different levels of stress depending on layer thickness. This result suggests that cavity growth may occur above a critical level of stored energy rather than above a critical stress. If the adhesive layer is predominantly elastic as shown on figure 1, this stored elastic energy can be approximated by the work done by the applied stress during the initial tensile part of the stress-strain curve; namely by:

$$U_{el}(e) = h_0 \int_0^e s\, de' \quad (1)$$

The number of visible cavities per unit area can now be plotted as a function of $U_{el}(e)$ (figure 5b). The comparison between the 60 and 120 µm layers is particularly instructive: For an identical initial defect population, the first cavities appear when the same amount of elastic energy is stored in the layer and not for the same average stress. In the 240 µm layer on the other hand, most cavities appear when a critical stress (of the order of 0.4 MPa in our case) is attained. This critical stress, identical for the large defect experiments of figure 3a, is a material property. Admittedly the stress in the layer is modified locally by the appearance of new cavities, but if we restrict our analysis to the portion of the curve before the peak stress, the appearance of cavities do not significantly alter the linearity of the curve (figure 2). This implies that the amount of energy stored, or the average stress, in a portion of the layer where a cavity has yet to appear must be proportional to the total energy stored in the layer.

In summary for the thicker layer, cavities expand above a critical stress, while for thin layers, a criterion of a minimum stored energy in the layer must also be met.

*Theory.* – The growth of a cavity in an elastic rubber is a classical problem in solid mechanics which can be treated analytically if spherical symmetry is preserved. Neglecting surface energy, if a preexisting cavity in a rubber elastic material is submitted to a tensile hydrostatic stress, it will expand spontaneously when the stress reaches approximately the elastic modulus $E$ [10]. However in reality the cost of increasing surface area contributes to prevent expansion. If no chemical bonds are ruptured during the expansion, it is simply the surface tension $g$; however generally the energy necessary to fracture chemical bonds should also be included [21] and the critical energy release rate $G_c$ should be used in place of $g$.

We can express the relationship between the negative hydrostatic pressure $P$ and the ratio $l$ between the inflated and the original cavity radius as:

$$P = \frac{E}{6}\left(5 - \frac{4}{l} - \frac{1}{l^4}\right) + \frac{2G_c}{r} \quad (2)$$

where $l$ is the ratio $r/r_0$ between the actual and the initial cavity radius. This expression defines a critical initial cavity radius: $r_c = G_c/E$, below which the second term of equation (2) dominates and expansion only occurs if the applied negative pressure $P$ is above $2G_c/r$. We have purposely neglected here the atmospheric pressure assuming therefore a solid far from the free surfaces.

In our experiments, $G_c$ can be estimated as approximately equal to $g \sim 30$ mJ/m$^2$ (for our block copolymer based adhesive, most of the elastic modulus $E$ comes from entanglements and the density of actual crosslink points provided by bridging triblock chains remains much lower). Since $E \sim 0.1$ MPa, $r_c$ is of the order of 300 nm. In other words any cavity of an initial size below 300 nm should expand at a level of stress, independent of the thickness of the layer (as predicted by equation (2)) in contradiction with figure 3c and also with some previously reported results [10,22].

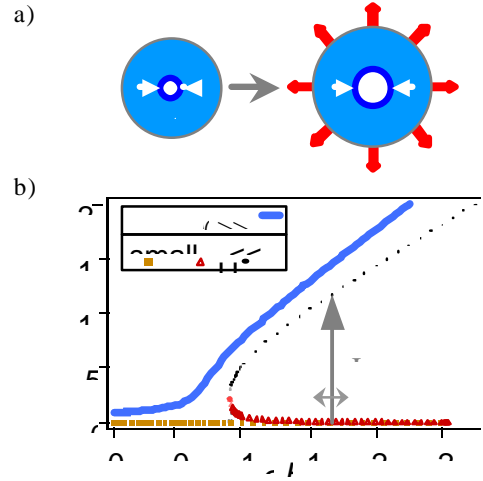

Figure 6: a) Schematic of the model growth of a cavity from an existing defect. b) Solutions minimizing the potential energy of the system when a hydrostatic tensile stress is applied to a preexisting cavity. The unique stable solution for the large cavity is shown in blue. For a small cavity, the metastable solution (M) appears in yellow, the unstable solution (U) in red and the stable solution (S) in black.

In order to explain our data, equation (2), which is written as a relationship between stress and strain, must be recast in terms of energy: one can compute the potential energy of the system and find its extrema which will correspond to stable, metastable or unstable equilibrium solutions [23]. Remaining in a spherically symmetric system (figure 6a), the important parameters that can be varied are $r_0$, $G_c$ and $E$ in addition to the total volume of the system [24]. The minima of this potential energy function can be calculated for a given ratio $G_c/E$, for different levels of applied hydrostatic stress and their position, $r(s/E)$ are represented on figure 6b for two particular values of $r_0$. The most interesting case is that of cavities with an initial radius $r_0$ smaller than $G_c/E$. For low values of applied stress (well below $E$), only one stable solution exists: the cavity does not expand. However for $E < s < G_c/r_0$, two solutions coexist and a transition from the metastable solution at $r \sim r_0$ to the stable expanded solution at $r >> r_0$ must occur as shown on figure 6b. The probability for this transition to occur will depend on the available elastic energy in a certain volume (proportional to $U_{el}$ in our experiments), $r_0$ and thermal fluctuations. For thick layers this probability is nearly one when the critical stress is reached, while for thin layers higher stresses are required to reach a measurable probability to overcome the barrier between the metastable state M (orange squares) and the stable state S (black dots).

This physical picture is now perfectly consistent with our results: the spontaneous expansion observed on figure 4b is actually a transition from one equilibrium state to another and the probability for this expansion to occur depends on the amount of available elastic energy, i.e. on the thickness of the layer.

The presence of an activation energy for nucleation is a classical feature of many processes. It has been, in particular, proposed by several authors [25-28] to explain the nucleation of voids in rubber particles in rubber-toughened polymers. However in that situation, no pre existing cavity is likely to be present, the cavitation stresses reach 20-30 MPa and all of these models cannot neglect the hydrostatic strain energy stored in the rubber particle, which can be neglected in our experiments.

To summarize there are three important and new results in our study:

- The experiments that we describe do not require the existence of a nucleation process, which is always somewhat mysterious mechanistically, but simply the transition from an energy activated growth for small defects to an energy-density (in other words stress) controlled growth with increasing initial cavity size. The existence of metastable solution in the potential energy (which will depend on the details of the elastic potential function of the rubber) is a necessary condition to have an energy-activated growth and is a key difference between cavitation in solids and cavitation in liquids.

- In the energy-activated growth, the volume of the material from which elastic energy is drained appears to be very large: the thickness of the layer where energy-activated growth is observed is probably at least 1000 times larger than the initial cavity radius. This value of 1000 is surprisingly large but highly reproducible and has been reported (without interpretation) by others [10,22].

- the critical defect size necessary to pass from one regime to the other, is dependent on the elastic modulus of the material. Since there is no reason to believe that the size of the defects present in the material depend much on their elastic modulus, the same distribution of defect sizes could lead to a stress controlled cavity expansion in high modulus rubbers and an energy-activated expansion in low modulus polymeric gels.

Clearly this energy activated mechanism of cavity growth cannot be valid indefinitely since cavities in a material with a vanishingly small elastic modulus such as a highly swollen elastic gel would then never grow. For those very soft materials however, controlled experiments of cavity growth are difficult and to the knowledge of the authors have not been yet performed. Our results suggest however that some interesting new regimes of cavitation may be observed. The increasing interest of physicists and material scientists in the mechanical behaviour of living tissues and cells should foster forthcoming research in this area since many biological organisms are indeed complicated macromolecular structures, highly swollen with water, but displaying some elasticity.

In conclusion, we have shown that the failure of a class of soft solids ($0.01 < E < 1$ MPa) under hydrostatic tension occurs by cavitation on defects. The criterion triggering the growth of these cavities depends essentially on the ratio between two length scales: the system size and the defect size and a material parameter $G_c/E$ representing the ratio between surface energy and elastic modulus.

For materials with preexisting cavities larger than the ratio $G_c/E$, spontaneous expansion of these cavities occurs above a critical stress ($s > s_{cav}$). For materials with defects smaller than the ratio $G_c/E$, the spontaneous growth process depends on sample size: for samples larger than ~1000 $G_c/E$ the critical stress criterion applies; for smaller samples, cavity growth becomes an energy activated process which depends on the total elastic energy stored in a much larger volume than previously thought.

* To whom correspondence should be addressed.
Electronic address: Costantino.Creton@espci.fr